\documentclass[aps, twocolumn, showpacs]{revtex4-1}
\usepackage{amsmath}
\begin{document}
\title{Proton spin: a topological invariant}
\author{S. C. Tiwari \\
Department of Physics, Institute of Science, Banaras Hindu University, Varanasi 221005, \\ Institute of Natural Philosophy \\
Varanasi India\\}
\begin{abstract}
Proton spin problem is given a new perspective with the proposition that spin is a topological invariant represented by a de Rham 3-period. The idea is developed generalizing  Finkelstein-Rubinstein theory for Skyrmions/kinks to topological defects, and using nonabelian de Rham theorems. Two kinds of de Rham theorems are discussed applicable to matrix valued differential forms, and traces. Physical and mathematical interpretations of de Rham periods are presented. It is suggested that Wilson lines and loop operators probe the local properties of the topology, and spin as a topological invariant in pDIS measurements could appear with any value from 0 to $\frac{\hbar}{2}$, i. e. proton spin decomposition has no meaning in this approach.
 
\end{abstract}
\maketitle

\section{\bf Introduction}
The theoretical understanding on the origin of the proton spin was challenged by the famous EMC (European Muon Collaboration) experiment more than 25 years ago \cite{1}. The extensive reviews dedicated to proton spin problem \cite{2,3} show that in spite of many experiments since then, and also the advances made in the perturbative QCD and lattice QCD calculations the proton spin problem has not been solved satisfactorily. Leaving aside the so called proton spin decomposition controversy \cite{4} nicely reviewed by Wakamatsu \cite{5} and Leader and Lorce \cite{6} it may be asked if a new line of thinking would be fruitful. Scattering experiments have firmly established that proton is not a point particle, and it has a 3-dimensional spatial internal structure. It would be natural to envisage purely geometric and topological aspects of the internal structure to characterize physical properties of the proton. The present paper has a modest aim in this direction: to offer a tangible topological framework to understand proton spin as a topological invariant.

DIS experiments and the Bjorken scaling paved the path for the development of QCD as a unique quantum field theory (QFT) for the strong interaction; see textbooks \cite{7,8,9}. Though color confinement is a logical hypothesis within QCD it is somewhat unsatisfactory that the fundamental constituents of the hadrons, i. e. quarks and gluons, do not have physical reality as free particles. As regards to QCD topology it is extremely nontrivial with different facets being studied by the QCD physicists. The role of topological structure of the QCD vacuum in the proton spin problem is even more elusive \cite{2,3}. It is believed that, if any, its contribution to proton spin would be insignificant; however it may not be so as argued in \cite{10}.

There exists an alternative idea inspired by the nontrivially rich topological structure in the nonlinear field theories. In particular, in the low-energy limit of QCD the effective Lagrangian for chiral $SU(2)\otimes SU(2)$ and $SU(3)\otimes SU(3)$ have attracted a great deal of attention since the discovery of topological invariant by Skyrme \cite{11}. The anticipations on the unusual spin and statistics connection of the kinks/solitons by Finkelstein and Rubinstein \cite{12} played a key role in further developments. We shall refer this work as FR theory in the text. Wess-Zumino-Witten (WZW) topological invariant \cite{13,14,15} indicates a plausible interpretation of baryons as Skyrmions. It has been asserted \cite{15,16} that WZW term does lead to octet spin 1/2 and decuplet spin 3/2 baryons. Though these works indicate spin-topolgy connection they hardly figure in the literature on proton spin problem; a passing notice is taken by Bass in \cite{2}. One reason, of course, seems to be the speculative nature of this idea. More importantly it could be due to the lack of linking topology with the experiments: the experimental measurements on the structure functions in high energy scattering experiments have to be understood in terms of the topological structure.

To make progress in this direction a novel conceptual framework may be required. For this purpose first we make following points.

{\bf I.} Enormous literature beginning with Witten's papers \cite{14,15} has given significance to mainly one important aspect of FR theory, namely, that solitons could be fermions. However Section III in \cite{12} on spin introduces crucial distinction between extrinsic/intrinsic and otbital/spin angular momentum applicable to kinks and point field theories respectively. Further FR theory assumes continuity of the fields in the spacetime, in fact, the authors make it clear that without going into the dynamics of the fields, boundary conditions and the continuity are sufficient to arrive at their results. Now topological objects may have intrinsic discontinuities or defects, in that case FR theory would not be directly applicable. In the context of angular momentum we propose to generalize FR theory.

{\bf II.} Homotopy theory is widely used in the study of topology in physics; FR theory and Witten's work also make extensive use of it. However in some cases cohomology  groups may be more convenient than homotopy \cite{17}. To define cohomology groups one may proceed with a complex: the complex of differential forms defines de Rham cohomology group $H^n(X;R)$, where X is a differentiable manifold. It is very pleasing that de Rham cohomology perspective has been given to WZW term by D'Hoker and Weinberg \cite{18}. We argue that de Rham periods have utility in representing topological defects; we have earlier suggested their application in a topological model of photon \cite{19} and also discussed them in the proton spin problem \cite{10}. Unfortunately de Rham periods are not quite familiar in the modern discussion on de Rham cohomology though standard text \cite{20} discusses them. The classic of Hodge \cite{21} and the articles \cite{22,23} are referred for details.

{\bf III.} Parton model and hard-scattering factorization theorems in QCD have been immensely powerful tools in the high energy hadron physics \cite{8}. Unfortunately the same cannot be said for nonperturbative QCD. There are numerous models for color confinement; quite a few being exotic. In a class of models monopoles in QCD play a key role in the mechanism of color confinement, for example, see \cite{24,25,26}. It seems Schwinger in one of early papers put forward a highly speculative idea that hadrons are composite particles of dyons \cite{27}. Though today the term monopole has acquired varied and sometimes weird facets, Wu and Yang \cite{28} cautioned that Dirac magnetic monopole \cite{29} is different than the nonabelian monopole.

Monopole problem goes into the Witten's construction for WZW term \cite{14}. In his second paper \cite{15} it is pointed out that the SU(2) soliton embedded in SU(3) is invariant under spin, isospin and hypercharge transformations; in the SU(3)/U(1) manifold the Wess-Zumino term could be interpreted as Dirac monopole-type, and the lowest monopole harmonics could be identified with SU(3) octet spin 1/2 and decuplet spin 3/2. Aitchison \cite{30} gives a nice discussion that WZW term in field space induces monopole-like effect in real space. All the heuristic arguments suggest monopole and Dirac quantization implying spin half. In the literature there exist two conceptually distinct approaches: Dirac quantization $\rightarrow$ spin-half, and spin-half $\rightarrow$ Dirac quantization. We may state it in the form of monopole-spin equivalence, and in view of the nature of monopole being a topological defect it could be argued that spin too has a topological origin.

{\bf IV.} Polarized DIS experiments study the interior of the proton; if spin is a topological invariant the most important question becomes as to how the local structures of the topological objects are probed. A clue is provided by Aharonov-Bohm effect. Section 4 of the original paper \cite{31} considers the problem of electron beam scattering from a magnetic flux confined to a finite region of space, and it is shown that the scattering cross section has imprints of the local behavior of the topological object taking the point limit for the flux region. Wilson lines would be the ideal objects to probe the local aspects of proton spin.

The paper is organized as follows. Topological perspectives on the proton spin problem are reviewed in the next section. A brief discussion on the salient features of FR theory and its generalization for topological defects constitute Section III. In Section IV two kinds of nonabelian de Rham theorems applicable to matrix valued forms and traces of the forms are presented. Proton spin in the topological approach as a de Rham 3-period is discussed in Section V. Physical implications and the prospects of the proposed new approach constitute the last section.

\section{\bf Proton spin sum rules and topology}

{\bf A. Gluon Topology}

Spin structure functions $g_1(x, Q^2),~ g_2(x,Q^2)$ determine the proton spin, however the experimentally measured quantities and their interpretation depend on pQCD descriptions or models. The first moment of $g_1(x, Q^2)$ contains the isovector, octet and flavor-singlet axial charges; here Q denotes the invariant momentum transfer and x is the Bjorken variable. It seems the contribution of the second spin structure function is negligible \cite{3}
\begin{equation}
\int^1_0 g_2(x,Q^2) ~dx =0
\end{equation}
The important factor is the flavor-singlet axial charge $g^{(0)}_A$ that depends on the renormalization group factor, and also on the estimates of the octet axial charge. Formally one may write the QCD value as
\begin{equation}
g^{(0)}_A =\sum _q \Delta q -\frac{3\alpha_s \Delta g}{2\pi} +C_\infty
\end{equation}
The first two terms represent the quark and polarized gluon spin contributions; see Eq.(10) in \cite{2}. The term $C_\infty$ vanishes in pQCD, and represents the nonperturbative QCD vacuum effect.

Unfortunately the calculation of $C_\infty$ depends on the nontrivial QCD vacuum structure, and its measurement in pDIS experiments is hardly possible as the nonlocal gluon topology at x=0 is believed to be important. However the role of topological current in $C_\infty$ is worth mentioning. Let us recall that the proton spin vector is defined in terms of the forward matrix element of the proton axial-vector current $\bar{\psi_p} \gamma_\mu \gamma_5 \psi_p$, and the quark spin contributions to this are obtained through the quark axial charges obtained from quark axial-vector current $\bar{q} \gamma_\mu \gamma_5 q$ for each flavor. The inevitably present QCD anomaly affects $g^{(0)}_A$; pQCD axial anomaly gives $\Delta g$ in the expression (2). Jaffe and Manohar \cite{32} argue that QCD anomaly has a nonlocal gluonic part because of the large gauge transformations of the gluon Chern-Simons current
\begin{equation}
K_\mu =\frac{g^2}{32 \pi^2} \epsilon_{\mu\nu\rho\sigma}[A^\nu_a(\partial^\rho A^\sigma_a -\frac{1}{3} g f_{abc} A^\rho_b A^\sigma_c)]
\end{equation}
Of course, once a specific gauge is chosen i. e. the 
light-cone gauge the anomaly current $K_+$ gives just the gluonic spin contribution. Here the light-front coordinate description is defined by $x^\pm =(t\pm z)/\sqrt 2$ and $x^\mu=(x^+, x^-, {\bf x}_t)$. And the light-cone gauge is defined by $A_+=0$. What happens in any other gauge, e. g. the covariant gauge? It is clear that the role of gauge invariance in the axial anomaly and the nature of QCD vacuum are crucial to understand this topological term.

{\bf B.  Why Topology?}

The 'small' gauge transformation in pQCD does not reveal the topological structure of QCD vacuum. Large gauge transformations and the boundary conditions at infinity have to be carefully understood. Lowdon has analyzed boundary terms in the axiomatic formalism of QFT in which the fields are operator-valued distributions \cite{33}. He assumes local (anti-) commutativity axiom and modified weak Gauss law to establish the behavior of spatial boundary terms in QFT. Locality axiom physically corresponds to the causality or compatibility of individual measurements in the space-like separated regions \cite{34}. The second assumption ensures local field algebra for quantized gauge field theories. Lowdon arrives at the result that a necessary and sufficient condition for the boundary term operators to vanish is that the operators annihilate the vacuum.

At a fundamental level the questions are those of mathematical continuum and physical reality of point particles. Mathematically the continuity to a spacetime point is formally approached through a logically consistent procedure of distribution theory. Though the infinitesimal calculus becomes operationally applicable the geometrical and topological properties of the spacetime manifold are lost in the process. Further on physical grounds, point particle is not the physical reality but an approximation to it. Lowdon also recognizes the limitations of a spacetime point in QFT in the introductory part of his paper, i. e. quantum measurement is not well-defined. Rather a measurement is performed over a finite extended spacetime region and the physical quantity is determined by an operator as a distribution smeared with some spacetime test function. The choice of the test functions and their support is arbitrary, therefore, a topological structure cannot be probed. Another consequence of the operator-valued distributions is that the boundary terms are also not well-defined in the integrations.

Author makes application to the proton spin problem making QCD angular momentum decomposition following Belinfante procedure, and it is argued that the total divergence term should not be dropped. The total angular momentum tensor $J^{\mu\nu\lambda}_{QCD}$ has the angular momentum charge $J^i_{QCD}$ decomposition as follows
\begin{equation}
J^i_{QCD} =L^i_q +S^i_q +L^i_g +S^i_g +(S^i_1 +S^i_2)
\end{equation}
The last two terms in this expression are the surface terms. If one assumes vanishing surface terms the matrix elements between z-polarized proton state gives the proton spin sum rule
\begin{equation}
\frac{1}{2} = \frac{1}{2} \Sigma +L_q +S_g+ L_g
\end{equation}
Lowdon demonstrates two important results: 1) the surface terms do not necessarily vanish, and 2) more surprisingly the matrix elements of $S^i_1$ and $S^i_2$ cancel those of quark spin $S^i_q$ and gluon spin $S^i_g$ respectively. Though kinetic versus canonical angular momentum is a debatable issue \cite{5,6} it is not of concern here. The significance of these results as noted by Lowdon himself is that nonperturbative QCD has important role in the proton spin sum rules. Could it come from topology?

There is another reason to take seriously the idea that spin has origin in topolgy. Let us consider SU(3) flavor symmetry group classification of hadrons. For mesons as composite of quark-antiquark pairs the total angular momentum is assumed to be a vector sum of the spin of individual quarks and their relative orbital angular momentum. The role of $S_g$ and $L_g$, in principle, could not be restricted to the proton spin. It may be asked if there is no inner angular momentum structure of mesons. Though not much work has been done on this question, relatively recently some interesting studies have been reported in the literature. A nontrivial transverse spin structure of the pion is indicated in the lattice QCD calculations \cite{35}. Authors suggest that the Boer-Mulder function for transverse polarization of quarks in an unpolarized hadron seems to be similar for valence quarks in both pions and nucleons. It may be recalled that the nucleon spin in the naive model is assumed to be that of three valence quarks in the s-orbital state. Now SU(3) isospin multiplets for baryons, for example, $J^P=\frac{1}{2}^+$ octet and $J^P=\frac{3}{2}^+$ decuplet have fixed spin irrespective of the nature of the constituent quarks. The universality of spin for a hadron multiplet does not have a simple explanation; it seems logically appealing to treat spin as a topological invariant for a multiplet.

{\bf C.  Differential Geometric  Approach}

An alternative idea has been suggested \cite{10} to understand the proton spin decomposition. The main point is that a topological defect could be represented by a nonvanishing de Rham period, and the standard total angular momentum decomposition is modified with a topological term ${\bf J}_{top}$. Thus the total angular momentum
\begin{equation}
{\bf J} = {\bf S}_q+{\bf S}_g+{\bf L}_q+{\bf L}_g 
\end{equation}
gets modified to
\begin{equation}
{\bf J}_{mod} = {\bf J} + {\bf J}_{top}
\end{equation}
The abstract discussion on de Rham theorems in QCD given in \cite{10} seems to be one of the reasons that this idea did not attract much attention. Another, perhaps more important reason is that unlike homotopy theory de Rham theorems are not so familiar among QCD physicists. We present salient features of this approach in Section IV. In the present paper the most profound new insight is the proposition that the entire proton spin has topological origin; it would mean replacing expression (7) by the following
\begin{equation}
\frac{1}{2} = {\bf J}_{top}
\end{equation}

\section{\bf Skyrmions/kinks and topological defects}

{\bf A.  Kinks and  Spin}

The significance of spin-statistics relation derived in the multivalued quantization of FR topological approach has been widely noticed in the literature. We highlight certain additional insights in this paper \cite{12}. A nonlinear classical field theory could develop indestructible spatially extended objects called "kinks". The assumption on continuity is sufficient to establish spin-statistics connection based on homotopy calculations; there is no need for detailed dynamics of the system. The topological approach to kinks makes it necessary to adopt what the authors call a refined terminology on spin. Half-odd spin is just a sign change under $2\pi$ rotation. For the angular momentum the decomposition is not spin and orbital parts but intrinsic and extrinsic parts
\begin{equation}
{\bf J} ={\bf J}^i +{\bf J}^e
\end{equation}
Since the angular momentum is a generator of infinitesimal rotations we consider the total change in the field to have two parts: extrinsic change due to spatial rotation and the intrinsic transformation of the field. The corresponding generators give the decomposition (9). If the physical field is represented by multivalued functionals then the extrinsic angular momentum may contribute to the spin change. Thus the standard decomposition
\begin{equation}
{\bf J} = {\bf L} + {\bf S}
\end{equation}
is unsatisfactory as orbital part may contribute to the spin of the kink. It would appear that the spinors in QFT are 'point-limit' approximation to the internal fields of the kinks. Since proton is not a point particle it is easy to appreciate the importance of this distinction between the two decompositions (9) and (10) in the context of proton spin problem.

In the homotopy approach the authors introduce a covering space where the multivalued function is single-valued, and discuss the action of flow that uniquely defines a path in the covering space. A closed flow has closed paths or loops. A flow for a continuous $2\pi$ extrinsic rotation is a loop in SO(3) manifold: 3-dimensional ball of radius $\pi$ with opposite points of the surface identified. The paths for the exchange of kinks and rotation loops are discussed using homotopy extension theorem: the existence of a continuous function on a manifold M with prescribed boundary conditions  on the boundary $\partial M$. For their purpose a continuous function $\phi({\bf x},s,t)$ defined on $X \otimes I\otimes I$, where I is unit interval and X is ordinary Euclidean space, is considered. One may, in fact, use a 3-cube for X thereby the function $\phi({\bf x},s,t)$ is defined on 5-cube $I^5 = I^3 \otimes I \otimes I$. Though $\partial I^5$ is 4-dimensional only z=constant cross sections are sufficient. Thus the construction used is that of $I^5|_{z=const}$ a 4-cube $I^4$,    a 3-dimensional manifold homeomorphic to the 3-sphere $S^3$. Table 1 in \cite{12} lists the homotopy groups for the classical Lie groups.

Authors admit that taking point-limit of kinks has many technical problems; note that the kinks are continuous extended structures. How does spin arise from orbital motion? A nice picture is suggested in which rotating asymmetric field distribution may contribute to the spin part. The important conclusion is that the FR theory as such is not applicable to intrinsic field singularities giving topological defects.

{\bf B.  Spin of Topological Defects}

The difference between a kink and a defect is best understood considering the example of a monopole. Dirac magnetic monopole has string or line singularity that can be removed by Wu-Yang construction or other topological method, however at the origin the point singularity is real and cannot be removed. Thus Dirac monopole is a defect, not a kink. On the other hand the nonabelian monopoles, for example, 't Hooft-Polyakov monopole and $SU(2) \otimes SU(2) ~ \sigma-$model monopole \cite{25} are kinks/solitons. In these monopoles a non-singular field configuration and boundary conditions determine the topology, and the spin may have orbital contribution as suggested in FR theory. For a Dirac monopole the spin and Dirac quantization have origin in the singularity; we may put it in the form of a nonvanishing de Rham period similar to the semiclassical Bohr-Sommerfeld quantization in Einstein's construction of a 2-torus orbital manifold \cite{36}.

The expression (9) of FR theory has to be generalized to describe the topology of defects. To develop the logic of our proposition we begin noting the well-known Casimir invariants for the inhomogeneous Lorentz group
\begin{equation}
p^\mu p_\mu =m^2
\end{equation}
\begin{equation}
W^\mu W_\mu =-m^2 s (s+1)
\end{equation}
\begin{equation}
W_\mu =-\frac{1}{2} \epsilon_{\mu\nu\rho\sigma} J^{\nu\rho} p^\sigma
\end{equation}
In the point field theory the interpretation is that mass and spin are characterized by the Casimir invariants. It is not necessary here to go into the technical details for massless particles and QFT. Obviously for kinks/defects this approach is not applicable. Even the usual expression (10) is not generally true for kinks as discussed in FR theory.

In the Lagrangian formulation of field theory the angular momentum density is a third rank tensor obtained as a Noether current for continuous spacetime symmetry; in QCD we have $J^{\mu\nu\lambda}_{QCD}$ and the angular momentum charge is
\begin{equation}
J^i_{QCD} =\frac{1}{2} \epsilon^{ijk} \int J^{0jk}_{QCD} d^3x
\end{equation}
Note that the SU(3) group index does not appear in the angular momentum tensor. The Lagrangian density also does not have SU(3) index, for example, the gluon field strength tensor
\begin{equation}
G^a_{\mu\nu} = \partial_\mu A^a_\nu - \partial_\nu A^a_\mu - g f_{abc} A^b_\mu A^c_\nu
\end{equation}
gives the following Lagrangian density 
\begin{equation}
\mathcal{L}_{QCD} = -\frac{1}{4} G^a_{\mu\nu} G^{a,\mu\nu}
\end{equation}
Since the trace is just a number, we have two kinds of objects in the nonabelian gauge theories: matrix-valued quantities and traces which lead to the difference in the corresponding de Rham theorems as we will see in the following.

Let us now consider a field configuration $\Psi$ that has finite spatial extension like a soliton, and an intrinsic singularity then the proposed generalized total angular momentum is defined to be
\begin{equation}
{\bf J}_{gen} = {\bf J}_e +{\bf J}_i +{\bf J}_{top}
\end{equation}
For a field with only singularity it becomes
\begin{equation}
{\bf J}_s ={\bf L} +{\bf J}_{top}
\end{equation}
The intrinsic topological invariant is defined to be the spin
\begin{equation}
{\bf S} = {\bf J}_{top}
\end{equation}
To understand the difference between the intrinsic angular momentum and spin note that ${\bf J}_i$ originates from the transformation law of the field $\Psi$ under rotation, i. e. a vector or a spinor whereas spin ${\bf S}$ is proposed to be a purely topological effect due to the presence of the field singularities or topological obstructions. The property (19) is not dependent on the nonlinearity of the field equations, and cannot be changed by orbital rotation.

Where does lie the topological information on spin? Physically the universal spin of hadron multiplets would correspond to this. The mathematical object is suggested to be a nonvanishing de Rham period: if Bohr-Sommerfeld quantization is followed the spin is represented by a 1-period, while assuming it to be a generalization of the third rank angular momentum density tensor the spin is a 3-period. The quantization rules for the periods with the input from physics that these are integral multiples of some basic unit give the spin quantization assuming the unit to be $\hbar$. Half-integral spin can arise only if the oppositely oriented boundaries of the cycle exist.

\section{\bf Nonabelian de Rham Theorems}

The utility of the language of differential forms is recognized in the modern texts on QFT; the term 'de Rham period' is however, not very familiar, and sometimes doubts are expressed as to whether de Rham theorems are applicable for a Lorentzian spacetime manifold and nonabelian gauge theories. A lucid treatment is given by Nash \cite{17}; see also \cite{4}. For the sake of clarity and completeness a brief account is presented here so that the significance of the two kinds of de Rham theorems in nonabelian  case becomes clear.

The well-known examples of two dimensional surface integral of Gauss and one-dimensional loop integral of Ampere in electromagnetism probe the interior of the integration domains. Their extension to higher dimensional manifolds is referred to as generalized Stokes theorem or formula: exterior differential forms and the integration over chains in a given manifold M of dimension n formally define the Stokes theorem
\begin{equation}
\int_c d\omega =\int_{\partial c} \omega
\end{equation}
where the domain of integration c has the boundary $\partial c$, and $d\omega$ is the exterior differential of the differential form $\omega$. The formal structure of (20) has the profound implication: the topology of the integration domains (homology) and the topology of the differential forms (cohomology) are dual to each other. Metric-independence of (20) implies that this formula is not restricted to Riemannian metric manifold: it can be seen explicitly \cite{17} considering the definition of a p-form (or a form of degree p). A totally anti-symmetric covariant p-tensor field using local coordinates and the exterior or wedge product defines the p-form $\omega$
\begin{equation}
\omega =\omega_{i_1 ....i_p}~ dx^{i_1}\wedge dx^{i_2} ...dx^{i_p}
\end{equation}
The Hodge star operator defines a unique dual (n-p)-form
\begin{equation}
^*\omega = \frac{\sqrt g}{{(n-p)}!} \epsilon^{i_1...i_p}_{i_{p+1}...i_n} ~dx^{i_{p+1}} \wedge ...dx^{i_n} 
\end{equation}
Here g is the determinant of the metric tensor of the Riemannian manifold. The volume n-form $\tau$ defines the Hodge dual such that for every $\alpha$ belonging to the space of p-forms $\Lambda^p(M)$ the inner product is
\begin{equation}
\tau (\alpha|\omega) =\alpha \wedge *\omega
\end{equation}
For a Lorentzian manifold: the determinant is $\sqrt{-g}$ and the Hodge star is
\begin{equation}
*^2 =-1{(-1)}^{p(n-p)}
\end{equation}

In the Stokes formula the chain $c$ and the boundary $\partial c$ are coherently oriented, and it is applicable for integrals of p-forms over p-chains. The set of differential forms $[\omega^i]$ and the set of chains $[c_i]$ constitute a graded vector space
\begin{equation}
d\omega^i \subset \omega^{i+1} ~ (d^2 =0)
\end{equation}
\begin{equation}
\partial c_i \subset c_{i+1}  ~ (\partial^2 =0)
\end{equation}
As a consequence to the Stokes theorem one has 
\begin{equation}
d^2 =0  ~\Rightarrow ~\partial^2 =0
\end{equation}
Now the usual statement of Poincare lemma is that
\begin{equation}
d(d\omega) =0
\end{equation}
or for a p-form $\omega$ on M and a (p-1)-form $\alpha$ such that $d\alpha =\omega$ then $d\omega=0$. In this case $\omega$ is called a closed form (or a cocycle). The converse of the lemma is that if $d\omega=0$ then $\omega =d\alpha$. This is not true globally: a form $\omega$ such that
\begin{equation}
\omega=d\alpha
\end{equation}
is called an exact form (or a coboundary). Similarly for the chains we define a cycle to be a chain $c$ such that $\partial c=0$, and a boundary such that $c=\partial B$. 
\begin{equation}
c=\partial B ~ \Rightarrow ~ \partial c =0
\end{equation}
but not necessarily
\begin{equation}
\partial c =0 ~ \Rightarrow ~ c=\partial B
\end{equation}
and 
\begin{equation}
\omega =d\alpha ~ \Rightarrow ~ d\omega =0
\end{equation}
but not necessarily
\begin{equation}
d\omega =0 ~\Rightarrow ~ \omega =d\alpha
\end{equation}
From (30)-(31) and (32)-(33) we state the following. The homology of M is the set of equivalence classes of cycles $Z_p$ of degree p that differ by boundaries $B_p$ i. e. $H_p(M) =Z_p/B_p$. On the other hand, de Rham cohomology $H^p(M)$ is defined as the equivalence set of closed differential forms $Z^p$ modulo the set of exact forms $B^p$ i. e. $H^p =Z^p/B^p$. The dimension of $H_p(H^p)$ is called the Betti number $b_p(b^p)$ of M. 

The global properties of a manifold are best understood using de Rham theorem that could be stated as the duality of the vector spaces $H_p$ and $H^p$, and $b_p=b^p$ for finite dimensional $H_p$ and $H^p$; it also establishes the duality of $H^p$ with compact support to the infinite chain homology for orientable manifolds \cite{20}.

An alternative form of this theorem is obtained defining a de Rham period. The integral of a closed form over a cycle is called a period. Obviously de Rham period depends on the homology class of the cycle and the cohomology class of the differential form. Thus we state the first and second de Rham theorems:

{\bf I. ~}For a given set of periods $[\nu_i]$ there exists a closed p-form such that
\begin{equation}
\nu_i = \oint_{c_i} \omega
\end{equation}
where $[c_i]$ is a set of independent cycles. The closed form is determined up to the addition of an exact form.

{\bf II. ~} If all the periods of a closed form vanish then it is exact.

To appreciate above theorems it may be worth reproducing the reformulated de Rham theorem in \cite{20}:

"There exists a closed form which has on $X^n$ arbitrarily preassigned periods on linearly independent homology classes. This closed form is determined up to the addition of an exact form." 

And, the corrolary: A closed form is exact if and only if all its period vanish.

Above background would help us to generalize de Rham theorems for the nonabelian theories. Let us note that in the topological approach Wilson loops and nonabelian Stokes theorem are fundamental geometric objects. Some technical details are necessary to develop the proposition on de Rham theorems in QCD made in \cite{10}. Following  \cite{37,38} we present the essentials.

Differential 1-form of the 4-vector gauge potential is denoted by
\begin{equation}
A=A^a_\mu t_a dx^\mu
\end{equation}
In \cite{38} the notation used is $A_\mu = gA^a_\mu t_a$. A continuous curve parameterized by s defines a path starting at $x_0$ and ending at x by $z^\mu (s=1) =x^\mu$ and $z^\mu (s=0) =x^\mu_0$, and a Wilson line appears in the solution of parallel transport equation
\begin{equation}
W(x, x_0) = P exp~[ig~\int^x_{x_0} ~\frac{dz^\mu(s)}{ds} A_\mu(z(s)) ds]
\end{equation}
Geometrically (35) is interpreted as connection 1-form. Since the path is divided into an infinite number of infinitesimal elements the matrix valued integrals have to preserve a path ordering P. Under a gauge
transformation Wilson line transforms as
\begin{equation}
W(x, y) ~\rightarrow ~ U(x) W(x, y)U^{-1}(y)
\end{equation}
The combination $\bar{\psi}(x)W(x, y) \psi (y)$ is gauge invariant; it is for this reason that a Wilson line is inserted between quark field operators at different positions. This gauge invariant combination depends
on the end points $(x, y)$ as well as the path that connects them. For light-like separation the Wilson line along $x^-$-axis depends only on the end points \cite{8}, and one can replace a direct line $W(x, y)$ by $W(x, +\infty)W(+\infty, y)$.

A Wilson loop is defined by a closed path such that $z_\mu(0)=z_\mu(1)$
\begin{equation}
W(L) = Tr P exp~[ig ~ \oint A]
\end{equation}
Stokes theorem relates the line integral with the surface integral over the surface enclosed by the loop. The nonabelian Stokes theorem \cite{37,38} is stated to be
\begin{equation}
P exp~[ig\oint_{\partial S} A] = P_S exp~[ig\int_S W(a, y) G(y) W(y, a)]
\end{equation}
The curvature 2-form is defined as
\begin{equation}
G = \frac{1}{2} G_{\mu\nu}^a dx^\mu \wedge dx^\nu =dA +A\wedge A
\end{equation}
On the rhs of (39) the symbol $P_S$ denotes surface ordering of infinitesimal area elements \cite{38}, $\partial S$ is the boundary of surface S, and a is a fixed reference point on the boundary. Introducing the notations
\begin{equation}
\bar{A}(y) =W(a, y) A(y) W(y, a)
\end{equation}
\begin{equation}
\bar{G}(y) = W(a, y) G(y) W(y, a)
\end{equation}
and using the properties of $W(a, y)$ the Poincare lemma is obtained
\begin{equation}
d\bar{A} =\bar{G}
\end{equation}
\begin{equation}
d\bar{G} =dd\bar{A} =0
\end{equation}
Here $\bar{G}$ and $d\bar{G}$ are gauge invariant apart from a position-independent gauge transformation at the reference point U(a), and (43) is gauge invariant and also independent of the reference point a. In the derivation of (44) use has been made of the Bianchi identity \cite{38}.

If there is no nontrivial topological structure, specially the absence of monopoles, then the nonabelian Stokes theorem is sufficient to define de Rham periods. However the interpretation of surface integral as a flux
is not unique since the field tensor $G_{\mu\nu}^a$ itself carries color. The presence of monopoles introduces additional complications: it has been shown that the QCD flux through a closed surface can be defined in the
loop space parameterized by two parameters s and t \cite{38}. The variation of parameter t from 0 to $2\pi$ gives a closed 2-dimensional space-like surface, and the closed surface $\Sigma$ obtained through loop
space replaces S on the rhs of (39) together with the change of G to $\bar{G}$. We define formally
\begin{equation}
\Theta (\Sigma) = P_tP_s exp [ig \oint_\Sigma \bar{G}]
\end{equation}
We have used symbolically the path orderings $P_tP_s$ as the Wilson line (36) appearing in the integral on the rhs of (45) has to be defined for each value of the loop space parameter t
\begin{equation}
W_{z_t} (s,0) =P_s exp~[ig \int^s_0 A_\mu(z_t(s)) \frac{dz^\mu_t(s)}{ds} ds]
\end{equation}
Chan et al \cite{39} define a connection in the loop space given by
\begin{equation}
\mathcal{A }=W_{z_t}^{-1} (s,0)G^a_{\mu\nu}(z_t(s))W_{z_t} (s,0)\frac{dz^\nu_t(s)}{ds}
\end{equation}
If the curvature corresponding to $\mathcal{A}$ vanishes there are no monopoles according to the interpretation in \cite{39}.

We propose two kinds of nonabelian de Rham theorems depending on whether the periods are matrix valued or traces. Marsh considers only matrix periods: 2-form $\Theta(\Sigma)$ defined by (45) is a de Rham period for a closed 2-form matrix; we denote the period for any closed 2-form matrix $\bar{\Omega}^2$ by
\begin{equation}
\Theta_2 = P_t exp~\oint_{c_2} \bar{\Omega}^2
\end{equation}
The first de Rham theorem is essentially the same stated earlier for the abelian case ({\bf I}) where now the closed form has to be taken matrix valued, and the second theorem gets modified to the following.

${{\bf II}_{mod}. ~}$

The closed 2-form matrix $\bar{\Omega}^2$ is exact if and only if all its periods give the identity matrix.

Marsh points out that at least for 2-forms one has unambiguous generalization of de Rham
theorems to the nonabelian case. In contrast to the abelian case in which all periods of an exact 
form vanish here we get the identity matrix.

The second kind of de Rham theorems are proposed for the de Rham periods of the trace of the matrix valued forms $\Omega$ over cycles
\begin{equation}
\mathcal{P} =\oint_C Tr \Omega
\end{equation}
The nonabelian (NA) de Rham theorems are stated as follows.

{\bf NA I.} Nonvanishing periods $\mathcal{P}$ determine the de Rham cohomology group.

{\bf NA II.} If all the periods of a closed form vanish then it is exact.

\section{\bf Proton Spin}

The preceding section establishes two kinds of de Rham theorems for nonabelian theories; and defines de Rham periods. Proton spin is proposed to be a topological invariant of a soliton like object having intrinsic defect: a field configuration explained in Section III B. The symmetry group is assumed to be SU(3), however no specific quark/gluon model is necessary. The proposition is that the spin has origin in a nonvanishing de Rham 3-period $\mathcal{P}_3$ vide de Rham theorem NA-I. It is important to realize that de Rham periods, though not commonplace in the literature terminology, are widespread in their usage with different names. We elaborate on two illustrative examples connected with the present discussion.

In the old quantum theory angular momentum quantization in Bohr's model was given a loop integral formulation by Sommerfeld \cite{40}. Ingenious geometrical construction introducing artificial boundaries to obtain closed curve is used by Sommerfeld to define a loop integral
\begin{equation}
\oint p~ dq = n h
\end{equation}
and both azimuthal and radial quantizations are assumed
\begin{equation}
\oint p_\theta ~ d\theta =n_1 h
\end{equation}
\begin{equation}
\oint p_r ~ dr = n_2 h
\end{equation}
Now momentum could be expressed as a gradient of a scalar ${\bf p} =-{\bf \nabla} S$, therefore, only multi valued function could be consistent with the nonvanishing loop integral (50). Post \cite{36} has pointed out that Einstein resolved the issue with the assumption of a 2-torus orbital manifold (orbifold) such that two-valued momentum field is single valued on internal azimuthal cycle (51) and/or  external meridional cycle (52). In the language of differential forms expression (50) is a nonvanishing de Rham 1-period.

The second example is that of a magnetic monopole. The electromagnetic field tensor is a closed 2-form F, and 2-period is $\oint_{S^2} F$. Noting that de Rham cohomology groups are $H^2(S^2;Z) =Z$ and $H^2(S^3;Z)=0$, F is closed but not exact on $S^2$ and it is exact on $S^3$. Dividing the 2-cycle $S^2$ into north and south hemispheres the de Rham period can be written in terms of the 1-forms $A_N$ and $A_S$ integrals along the oppositely oriented paths
\begin{equation}
\oint_{S^2} F= \oint_c A_N -\oint_c A_S
\end{equation}
Since $A_N$ and $A_S$ differ by a gauge transformation one can evaluate (53), and to obtain Dirac quantization condition the magnetic flux due to a monopole is used in the lhs of (53). It is of interest to know that monopole de Rham 2-period also relates with the Chern number. By definition the Chern class $C_m$ is the cohomology class of the closed 2m-forms \cite{20}. The integral of Chern class over a 2m-cycle is Chern number. In the monopole problem Chern number for the first Chern class is nothing but the de Rham period (53).

For the proton spin the suggested de Rham period is that of a closed 3-form over a 3-cycle. To calculate cohomology groups for SU(3) , it is noted \cite{17} that for this purpose there is no difference between SU(N) and U(N), and further that U(N) up to rational homotopy is equivalent to a product of odd dimensional spheres
\begin{equation}
U(N) ~\sim  S^1\otimes S^3.....\otimes S^{2N-1} 
\end{equation}
For $S^n$ the cohomology group is
\begin{equation}
H^m(S^n;Z) =Z ~if~ m=0~ or~ n;~ 0~ otherwise
\end{equation}
For SU(3) the nontrivial de Rham cohomology groups of interest are $H^3(SU(3);R)$ and $H^5(SU(3);R)$. Now de Rham theorem NA-I shows that the nonvanishing de Rham periods for these groups are $\mathcal{P}_3$ and $\mathcal{P}_5$. The arguments given earlier suggest that the natural object for spin would be $\mathcal{P}_3$: denoting the closed 3-form by $\Gamma^3$ we have
\begin{equation}
\mathcal{P}_3 =\oint_{C_3} \Gamma^3
\end{equation}
Though $\Gamma^3$ is not exact globally, similar to the monopole problem one may introduce two regions on the 3-cycle where the 3-form is locally exact
\begin{equation}
\oint_{C_3} \Gamma^3= \oint_{\partial C_3} \Gamma^2_+ -\oint_{\partial C_3} \Gamma^2_- 
\end{equation}

Note that by proper mathematical normalization one gets integers n for the periods, and for monopole like configurations one has n/2. Physical interpretation associates an elementary unit with the period, for example, the Aharonov-Bohm flux quantum $\frac{hc}{e}$ in the 1-period of the vector potential 
$\oint_{c_1} A = \frac{hc}{e} ~n$.
Assuming the elementary unit to be $\hbar$ for the 3-period, we state the proposition on proton spin as a topological invariant in the form
\begin{equation}
\mathcal{P}_3 =\oint_{C_3} \Gamma^3 =\frac{\hbar}{2} ~n
\end{equation}

In the introductory section (point III) we have discussed the significance of WZW term in the effective action for the spin of baryons and monopoles in the literature \cite{14,15,16,30}. D'Hoker and Weinberg \cite{18} have identified WZW term with the fifth cohomology group of SU(3). Using the matrix-valued 1-form $V=U^{-1} dU$
\begin{equation}
\Gamma^3 =\frac{i}{12\pi} Tr ~V^3
\end{equation}
\begin{equation}
\Gamma^5 =\frac{-i}{240 \pi^2} Tr ~ V^5
\end{equation}
Expression (60) corresponds to the WZW term, while (59) is related with the Goldstone-Wilczek topological current \cite{41}. Witten interprets this current determining the baryon number in a specific model \cite{14,15}. Goldstone and Wilczek \cite{41} give a different discussion on the physical significance of this current in particle physics. All the interpretations, however remain tentative and speculative. In our approach an alternative new significance is attached to $\Gamma^3$ given by expression (58).

\section{\bf Discussion and Conclusion}

A novel conceptual framework to understand proton spin has been presented in this paper. Two important questions arise immediately. What are the physical implications of the proposed abstract idea on proton spin as a topological invariant? What is the significance of the line of thinking presented here? Note that the motivation to explore this approach comes from the proton spin problem that has origin in pDIS experiments. The claimed new result (58) would be vacuous if it amounts to just a restatement that proton has spin-half. In marked contrast to the quark/gluon model in the present approach the issue of proton spin decomposition (6) itself becomes meaningless, and the measured value in the experiments represents information gained by the probes in pDIS experiments that can take any fraction k $(0\leq k\leq 1)$ of invariant proton spin $\frac{\hbar}{2}$. The inner geometrical structure is suggested to be described by Wilson lines that relate theory and experiment.

The logic behind above suggestion is traced to the Aharonov-Bohm effect: there is a vast literature on this subject and there are numerous ways to explain it. If it is considered as a topological effect then the original scattering theory given in \cite{31} could be viewed in terms of the Aharonov-Bohm phase as a probe of topology. To be precise, there are three objects of interest, namely phase, phase factor for open paths, and phase factor for closed paths given respectively by
\begin{equation}
\frac{e}{\hbar c} \int^b_a A_\mu dx^\mu
\end{equation}
\begin{equation}
exp ~\frac{ie}{\hbar c} \int^b_a A_\mu dx^\mu
\end{equation}
\begin{equation}
exp ~\frac{ie}{\hbar c} \oint A_\mu dx^\mu
\end{equation}
Wu and Yang \cite{28} explicitly recognize the different roles of the three, i. e.  (61)-(63). In the electron beam scattering by a magnetic field of finite region discussed in \cite{31} (61) and (62) can be used to determine the scattering cross section; however in the point-limit of the magnetic field region with flux remaining fixed the phase factor (63) gives information about the topological effect. Note that the topology enters into the picture for nonshrinkable loops in (63). Analogous to flux we have spin, and the Wilson lines (loops) carry the topological information on spin.

Two examples seem to support above arguments. The Sivers and Boer-Mulders functions defined without Wilson lines vanish as these are time-reversal odd. The role of the past-pointing or light-like in any other direction of Wilson lines depends on the process: Sivers function for SIDIS or TMD parton densities in the Drell-Yan process; see Section 13.17 in \cite{8}.

The second example is that of proton spin decomposition. 
Burkardt \cite{42} explained the difference between Ji and Jaffe-Manohar definitions of quark orbital angular momentum (OAM), and also gave a physical interpretation to the potential angular momentum suggested by Wakamatsu \cite{43}. In the present context we are interested only in the geometry of paths and Wilson lines. The ingenuity in Burkardt's
work lies in a U-shaped light-like path. In the light front coordinates $(0^-, {\bf 0}_t)$ is linked to $(\infty^-, {\bf0}_t)$ then to $(\infty^-,{\bf x}_t)$  and the path is completed returning to $(x^-,{\bf x}_t)$.
The Wilson line to achieve gauge invariance in the Wigner distributions is constructed of three straight line gauge links by Burkardt
\begin{align}
W_U^{+LC} &= W(0^-, {\bf 0}_t;\infty^-,{\bf 0}_t)~ W(\infty^-,{\bf 0}_t; \infty^-,{\bf x}_t) ~ \nonumber \\
     &\qquad W(\infty^-,{\bf x}_t; x^-, {\bf x}_t)
\end{align}.
In the LC gauge the gauge potential at light-cone infinity is pure gauge potential, and assumed to satisfy antisymmetric boundary conditions
\begin{equation}
{\bf A}_t(+\infty,{\bf x}_t) = -{\bf A}_t(-\infty, {\bf x}_t)
\end{equation}
Burkardt's aim is to calculate transverse momentum and OAM using two different paths: a direct straight line path from $(0^-,{\bf 0}_t)$ to $(x^-,{\bf x}_t)$, and a staple to $\pm \infty$, i. e. the U-shaped
path described above. From PT invariance it is argued that quark OAM calculated from $W^{+LC}_U$ and $W^{-LC}_U$ is equal, and the Jaffe-Manohar OAM can be taken equal to either of them. On the other hand Ji OAM
is obtained by a straight line path having 
\begin{equation}
W_{straight} = W(0^-,{\bf 0}_t; x^-, {\bf x}_t)
\end{equation}
The difference between the two is identified with the potential OAM.

We consider a closed path to define a Wilson loop; the first kind of de Rham theorems and matrix valued forms are useful here. The role of quark fields has to be viewed as some sort of auxiliary field representing the geometric structure. It is straightforward to see that a U-shaped path followed by a straight line path from $(x^-,{\bf x}_t)$ to $(0^-,{\bf 0}_t)$ results
into the desired closed path. The associated Wilson loop is given by
\begin{equation}
W_{QCD}(L) = W^{+LC}_U ~ W(x^-,{\bf x}_t: 0^-,{\bf 0}_t)
\end{equation}
Stokes theorem (45) shows that (67) leads to the nonabelian color flux enclosed by the loop, and the gauge field is given by
\begin{equation}
\bar{G} (x^-,{\bf x}_t) = W(0^-,{\bf 0}_t; x^-, {\bf x}_t) G(x^-,{\bf x}_t) W(x^-,{\bf x}_t; 0^-,{\bf 0}_t)
\end{equation}
Expression (68) offers the color flux interpretation of the potential OAM consistent with Burkardt's interpretation.
In the QCD the 2-form in (45) is $\bar{G}$, and the implication of second theorem is that a unique $\bar{A}$ does not exist. This arises because of the arbitrariness of the gauge rotation with position independent
U(a) on the gauge potential and the field tensor mentioned earlier. Wu and Yang \cite{28} as well as Fishbane et al \cite{37} have argued that nonabelian flux through a loop cannot be defined, however Marsh has clarified \cite{38}
that the arbitrariness due to path dependence is limited to position independent gauge transformations, and the nonabelian flux is a meaningful quantity.

To conclude prospects of the present approach for future development are outlined. The role of Wilson lines and loops becomes crucial for the envisaged nontrivial inner geometry of the proton: speculating that quarks/gluons are links/knots the color confinement acquires a different meaning as the impossibility of unknotting them. Hirayama et al \cite{44} have studied nonabelian Stokes theorem for nontrivial loops with interesting results that could be of use here. The scope of the present approach is not limited to proton, it has the potential to be developed for hadrons in a new perspective. The geometric analogue of quarks and associated flavors will have to be searched. A possible structure is that of 3-torus for three flavors: it is roughly a product of three circles having a 0-cycle, a 3-cycle, three 1-cycles and three 2-cycles. We plan to investigate if Wilson loops for 3-torus cycles could be related with the observed quantities. We understand that in the light of many interesting previous studies on vortices, monopoles and knots in QCD, developing the present adeas further would be a big challenge.

\end{document}